\documentstyle[12pt,epsf]{article}

\def\pd{\partial}
\def\a{\alpha}
\def\b{\beta}

\def\eps{\epsilon}

\def\e{{\rm e}}
\def\z{{\bar z}}
\def\w{{\bar w}}
\def\half{\frac{1}{2}}
\def\fr{\frac}
\def\T{{\hat T}}
\def\Tp{{\hat T}^+}
\def\Tm{{\hat T}^-}
\def\S{{\hat S}}
\def\pp{\prime}
\def\bb{\begin{equation}}
\def\ee{\end{equation}}
\def\bba{\begin{eqnarray}}
\def\eea{\end{eqnarray}}

\begin{document}

\begin{titlepage}

\begin{tabbing}
   qqqqqqqqqqqqqqqqqqqqqqqqqqqqqqqqqqqqqqqqqqqqqq 
   \= qqqqqqqqqqqqq  \kill 
         \>  {\sc KEK-TH-471} \\
         \>   hep-th/9602150 \\
         \>  { February 1996} 
\end{tabbing}
\vspace{5mm}

\begin{center}
{\Large{\bf $W_{\infty}$ Structures of 2D String Theory}}\footnote{
Talk given at the workshop on ``Frontiers in Quantum Field Theory'', Osaka, 
Japan, December 1995.}
\end{center}

\vspace{1cm}

\centering{\sc KEN-JI HAMADA}

\vspace{7mm}

\begin{center}
{\it National Laboratory for High Energy Physics (KEK)}, \\
{\it Tsukuba, Ibaraki 305, Japan}\\
hamada@theory.kek.jp
\end{center}

\vspace{1cm}

\abstract{
The Ward identities of the $W_{\infty}$ symmetry in 2D string theory 
in the tachyon background are studied in the continuum approach. 
Comparing the solutions with the matrix model results,  
it is verified that 2D string amplitudes are different from 
the matrix model amplitudes only by the external leg factors even in 
higher genus. 
This talk is based on the recent work [1] and also [2] for the 
$c_M <1$ model.}

\end{titlepage}

\section{Introduction}
   Many interesting issues in string theory such as dynamical  
compactification, black hole physics, etc,  require a non-perturbative 
formulation. At present, such a formulation is not available in higher 
dimensional string theories. In two or fewer spacetime dimensions, 
however, the string theory becomes solvable~\cite{k} 
due to the presence of   
$W_{\infty}$ symmetry~\cite{w,k2,fkn2,hh,h,hop,dmw},  
which gives a possibility of studying the non-perturbative formulation 
of string theory. Furthermore 2D string theory itself has interesting  
spacetime physics. It gives the 2D dilaton gravity coupled to a 
massless field called ``tachyon'' as the effective theory~\cite{np}:
\bb
   I_{\rm eff}= \int d^2 x \mbox{$\sqrt{-G}$}\e^{-2\Phi} \biggl[ 
        \fr{1}{4} R^G + (\nabla \Phi )^2 + 4  
           -\half (\nabla T)^2 +2 T^2 + \fr{2\sqrt 2}{3}T^3 
               \biggr] ~. 
\ee
Thus 2D string theory is also attractive as an alternative approach 
to studying 2D quantum dilaton gravity~\cite{h2}. 
 
  There are several formulations of two dimensional string theory.  
The matrix model (see reviews [3] and references therein)   
is generally believed to describe the 2D string 
theory, which is in principle defined non-perturbatively. 
The continuum theory~\cite{dk,s,p} 
is defined using the standard quantization 
method of the string perturbation theory. 
The topological description of 2D string theory is formulated 
in refs.[15,7]. 
To understand the non-perturbative formulation of string theory 
it is important to clarify the relations between these methods.  

  In the present work we investigate the continuum method of 2D string 
theory and show that the following relation is exact even in higher 
genus:
\bb
   S^{(g)}_{k_1,\cdots, k_N \rightarrow p_1, \cdots, p_M}
    = \prod^N_{i=1} \fr{\Gamma(-k_i)}{\Gamma(k_i)} 
         \prod^M_{j=1} \fr{\Gamma(-p_j)}{\Gamma(p_j)}
       ~\S^{(g)}_{k_1,\cdots, k_N \rightarrow p_1, \cdots, p_M} ~, 
\ee
where $S^{(g)}$ stands for 2D string amplitude of genus $g$ and 
$\S^{(g)}$ is identified with the matrix model amplitude of genus $g$.

  The strategy how to show that the above relation is indeed satisfied 
is the following. We can directly calculate the $1 \rightarrow N$ sphere 
amplitudes in the continuum method so that we can check 
the relation easily~\cite{k}. 
But, it is very difficult to
calculate general sphere amplitudes and higher-genus ones  in the 
continuum method. 
We here consider the Ward identities of the $W_{\infty}$ symmetry 
which give the recursion relations between amplitudes on different genus. 
We then compare the solutions with the matrix model results.

\section{Physical States and $W_{\infty}$ Currents}

  We consider 2D string theory in the linear dilaton background~\cite{p}, 
$G_{ij}=\delta_{ij}$ and $\Phi=2\phi$,  
which is the vacuum solution of 2D dilaton gravity (1), 
\bb
    I_0 = \fr{1}{4\pi}\int d^2 z \mbox{$\sqrt{\hat g}$}
            ({\hat g}^{\a \b} \pd_{\a} \phi \pd_{\b} \phi 
               + {\hat g}^{\a \b} \pd_{\a} X \pd_{\b} X 
               + 2 \phi {\hat R} ) ~, 
\ee
where $\phi$ is the Liouville field which is identified with the space 
coordinate and $X$ is the $c_M =1$ matter field which is identified with the 
(Euclidean) time. 

  A physical state with continuous momentum can only be the massless scalar 
called ``tachyon'' in the string terminology. In two dimensions the tachyon 
mode becomes massless in the linear dilaton vacuum. The tachyon vertex 
operator with momentum/energy $k (>0)$ is given by
\bb
     T^{\pm}_k =  \fr{1}{\pi} \int d^2 z 
         \e^{(2-k)\phi(z,\z) \pm ikX(z,\z)} ~, 
\ee
where $\pm$ denotes the chirality. 
The selection of $k>0$ is called the Seiberg condition~\cite{s}. 
We will postulate below that the amplitude including the anti-Seiberg 
$(k<0)$ states vanishes. 

  There is an infinite number of physical states at 
the integer momenta called the discrete states~\cite{bmp}, which      
are constructed from the OPE of the tachyon operators 
with integer momenta,
$
      V^-_{n+1}(z,\z)V^+_{m+1}(w,\w) 
           \sim ~\fr{1}{\vert z-w \vert^2}  \break
                R_{n,m}(w) {\bar R}_{n,m}(\w) 
$, 
where $n,m$ are positive integers~\footnote{We slightly change the 
notation of the subscript of $R$ and $B$ in refs.[1,2].} 
and $V^{\pm}_k (z,\z)$ is the exponential part of the tachyon 
operator (4).  The states $R_{n,m}$ are nothing but the remnants 
of the massive string modes in higher dimensional string theories. 
The discrete states $R_{n,m}$ form the 
chiral $W_{\infty}$ algebra~\cite{w}. We here normalize the fields 
such that
\bb
     R_{n,m}(z) R_{n^{\pp},m^{\pp}}(w) = \fr{1}{z-w}
       (n m^{\pp} - n^{\pp} m) R_{n+n^{\pp}-1,m+m^{\pp}-1}(w) ~.
\ee

   Besides these, at the same momenta, there are the BRST invariant 
operators with conformal dimension zero, $B_{n,m}$~\cite{bmp}, which 
satisfy the ring structure
$
    B_{n,m}(z) B_{n^{\pp},m^{\pp}}(w) = 
         B_{n+n^{\pp}-1,m+m^{\pp}-1}(w) 
$. 
Combining $R_{n,m}(z)$ and ${\bar B}_{n,m}(\z)$, we can construct the 
$W_{\infty}$ symmetry currents~\cite{w}
$
     W_{n,m}(z,\z) = R_{n,m}(z) {\bar B}_{n,m}(\z)  
$, 
which satisfy
\bb
     \pd_{\z} W_{n,m}(z,\z) 
         = \{ {\bar Q}_{BRST},[{\bar b}_{-1}, W_{n,m}(z,\z)] \}~, 
\ee 
where the algebra 
$\pd_{\z}={\bar L}_{-1}=\{ {\bar Q}_{BRST},{\bar b}_{-1} \}$ is used.

\section{Scattering Amplitudes of Tachyons}

  Let us consider the action in the tachyon background
\bb
          I= I_0 + \mu_B T_0 ~,
\ee
where $T_0 = \lim_{\eps \rightarrow 0} T^{\pm}_{\eps}$. 
The tachyon is massless so that $S$-matrix including $T_0$ vanishes. 
To ensure the non-decoupling of $\mu_B T_0$ we must make   
the bare tachyon background $\mu_B$ divergent as follows:  
$  \mu_B \rightarrow \fr{\mu}{\eps} $. 

  The $S$-matrix of tachyons in the tachyon background is defined by 
\bba
   &&  S^{(g)}_{k_1,\cdots, k_N \rightarrow p_1, \cdots, p_M} 
          = < \prod_{i=1}^N T^+_{k_i} \prod_{j=1}^M T^-_{p_j}>_g = \\  
   &&  \biggl( -\fr{\lambda}{2} \biggr)^{-\chi/2} 
            ~\delta \biggl(  \sum^N_{i=1} k_i - \sum^M_{j=1} p_j \biggr)  
                  ~\mu_B^s \fr{\Gamma(-s)}{2}   
               < \prod_{i=1}^N T^+_{k_i} \prod_{j=1}^M T^-_{p_j}~
                   (T_0)^s  >_g^{(free)} ~,
                      \nonumber
\eea
The superscript $free$ denotes the free field representation. 
The $\delta$-function and $\mu_B^s \fr{\Gamma(-s)}{2}$ come from the 
zero-mode integrals of $X$ and $\phi$ respectively~\cite{gl}.   
$g$ is the genus, $\chi =2-2g$ and $s$ is given by
$s= \sum_{i=1}^N k_i + \chi -N-M $. 

  The theory is tranlationally invariant in the time $X$,  
while is not in the space coordinate $\phi$.
So the factorization property of amplitudes are different from the 
usual string theory in the zero-dilaton vacuum $\Phi=0$. 
Let us introduce the eigenstate of the hamiltonian 
$H=L_0 +{\bar L}_0$~\footnote{$L_0$ is the zero-mode of the Virasoro 
generator including the ghost part.},
$| h,l;N>$ with the eigenvalue $\half h^2 +\half l^2 +2N$, where 
$|h,l;N=0>= {\bar c}c \exp [ (2+ih)\phi(0,0) +il X(0,0)]|0>$.    
The normalization is given by 
$<h^{\pp},l^{\pp};N^{\pp}|h,l,;N>_0 =-\fr{2}{\lambda} (2\pi)^2 
\break  
\delta(h^{\pp}+h) \delta(l^{\pp}+l) \delta_{N^{\pp},N}$. 
Note that the on-shell ($H=0$) state has purely imaginary $h$. 
$l$ must be real to preserve the translational invariance of $X$. 
The string propagator is given by $\fr{2}{H}$. 
So the factorization of 2D string 
ampitude into two parts is given in the form
\bba
 &&  < {\cal O}> 
        = -\fr{\lambda}{2} \sum^{\infty}_{N=0} 
               \int\fr{dh}{2\pi} \int\fr{dl}{2\pi}  
          < {\cal O}_1 |-h,-l;N > 
               \nonumber  \\
  && \qquad\qquad\qquad\qquad\qquad \times
            \fr{2}{\half h^2 +\half l^2 +2N}
           < h,l;N|{\cal O}_2 > + \cdots ~, 
\eea
where $\cdots$ denotes other channels. The zero-mode integral of $X$ 
ensures the conservation of energy so that $l$ is fixed, while the 
zero-mode integral of $\phi$ does not produce the $\delta$-function in 
the linear dilaton background. We then obtain the analytic function of 
$h$. So $h$ integral is non-trivial even in tree amplitudes. 
Naively we can deform the $h$ integral to the complex plane. It picks up 
the on-shell poles on the imaginary axis.

\section{Ward Identities of $W_{\infty}$ Symmetry}

   We introduce the normalized tachyon vertex operator 
\bb
     \T^{\pm}_k = \Lambda(k) T^{\pm}_k  ~,
               \qquad \Lambda(k) = \fr{\Gamma(k)}{\Gamma(-k)} 
\ee
and call the amplitude given by replacing $T^{\pm}_k $ 
in (8) with $\T^{\pm}_k$ the $\S$-matrix. 
Henceforth we consider the Ward identities in the form  
$  \fr{1}{\pi} \int d^2 z \pd_{\z} < W_{n,m}(z,\z) \break
 {\cal O} >_g = 0  $, 
where ${\cal O}$ is a product of the normalized tachyon 
operators. 

  Let us first calculate the operator product expansion (OPE) 
between the current and the tachyon operators, which is given in 
refs.[4,1], 
\bba
   &&   W_{n,m}(z,\z)~ \Tp_{k_1}(0,0)~ \Tp_{k_2} \cdots \Tp_{k_n} 
               \nonumber \\
   && \quad  = \fr{1}{z}~ n! ~\biggl( \prod^n_{i=1} k_i \biggr) 
                 ~ \Tp_{k_1 + \cdots +k_n -n +m}(0,0) ~,
\eea
where $\Tp_k (z,\z)$ is defined by replacing the integral in (4) with 
${\bar c}(\z)c(z)$. It was computed step by step from the $n=1$ formula 
to the general $n$.  This is analogous to the calculation of the OPE 
coefficients in CFT, where $\Tp_{k_j} ~(j=2,\cdots, n)$ just play a role 
of screening charges. 
Note that the OPE with the zero-momentum tachyon 
$T_0$ vanishes, but the OPE with the tachyon background $\mu_B T_0$ 
becomes finite due to the renormalization of $\mu_B$.

  The OPE with the tachyon $\Tm_p$ is easily calculated by changing the 
chirality. It is carried out by changing the sign of the field $X$ such 
that $\Tp \rightarrow \Tm$ and $W_{n,m} \rightarrow -W_{m,n}$ (the roles 
of $n$ and $m$ are interchanged). 

   The OPE singularity gives the linear term of the Ward identity.  
In addition we get the BRST-trivial correlator
$
      <\fr{1}{\pi} \int d^2 z \{ {\bar Q}_{BRST}, 
                    [{\bar b}_{-1}, W_{n,m}(z,\z)] \} \break
                                {\cal O} >_g  
$. 
Usually such a correlator would vanish. In this case, however, it gives 
the anomalous contributions from the boundary of moduli space.   
The boundary is described by using the string propagator in the form  
\bb
    D = \fr{1}{\pi} \int_{\e^{-\tau} \leq |z| \leq 1}
          \fr{d^2 z}{|z|^2}z^{L_0}\z^{{\bar L}_0} 
      =  \fr{2}{H} - \fr{2}{H} \e^{-\tau H}  ~, 
         \quad \tau \rightarrow \infty ~, 
\ee
where the second term of r.h.s. is the boundary.

  Let us first calculate the $n=1$ anomalous contribution. We then have 
to evaluate the following boundary contribution:
\bba
  && \lim_{\tau \rightarrow \infty} -\fr{\lambda}{2} \sum^{\infty}_{N=0} 
        \int \fr{dl}{2\pi} \int \fr{dh}{2\pi} 
            \int_{\e^{-\tau} \leq |z| \leq 1} d^2 z  
         < {\cal O}_1 [ {\bar b}_{-1}, W_{1,m}(z,\z) ] 
            \nonumber \\
  && \qquad\qquad\qquad  \times
         {\bar Q}_{BRST} \fr{-2}{H}\e^{-\tau H} 
           |-h,-l;N >< h,l;N| {\cal O}_2 >  ~.
\eea
We consider only the $N=0$ mode. As a result, the $N \neq 0$ 
contributions vanish exponentially as $\e^{-2N\tau}$. 
The $z$-dependence of the integrand is given in the form 
\bba
   &&[ {\bar b}_{-1}, W_{1,m}(z,\z) ]{\bar Q}_{BRST} 
       \fr{1}{H}\e^{-\tau H}  |-h,-l> = 
              \\ 
   &&  f(h,l)|z|^{\{ (m-1)(-ih-l+2)-2m\}} \e^{-\tau( h^2 +l^2)/2} 
     |-h+i(m-1),-l+m-1 > ~, 
             \nonumber
\eea 
where we use ${\bar Q}_{BRST}=\half {\bar c}_0 H + \cdots$. $f(h,l)$ is the 
calculable coefficient. Changing the variable to 
$|z|=\e^{-\tau x}$, where $0 \leq x \leq 1$, 
we get the following $\tau$-dependence: 
$2\pi \tau \exp [-\tau \{ \half (h^2 +l^2)+x (m-1)(-ih-l)\}]$. 
Thus the integrand 
is highly peaked in the limit $\tau \rightarrow \infty$. So we can exactly 
evaluate the integral of $h$ at the saddle point $h_{s.p.}=i(m-1)x$. 
We then get the expression
\bba
   &&\lambda \tau \sqrt{\fr{2\pi}{\tau}} \int \fr{dl}{2\pi} \int^1_0 dx 
         \exp \biggl[ -\fr{\tau}{2} \bigl\{ (m-1)x-l \bigr\}^2 \biggr] 
            f(h=i(m-1)x,l)  
              \nonumber \\
   && \qquad \times 
         < {\cal O}_1 | i(m-1)(1-x), m-1-l><i(m-1)x,l|{\cal O}_2 > ~.
\eea  
The $x$ integral is also evaluated at the saddle point 
$x_{s.p.}=\fr{l}{m-1}$ and produces the coefficient 
$\fr{1}{m-1} f(h=il,l)= \Lambda(m-1-l)\Lambda(l)$.   
We then get the boundary contribution
\bb
    \lambda \int^{m-1}_0 dl < {\cal O}_1 ~\Tp_{m-1-l}>
            < \Tp_l ~{\cal O}_2 > ~,   
\ee
where the $\Lambda$-factors are absorbed in the $T^+_{m-1-l}$ and $T^+_l$.  
The $l$ integral is restricted within the interval $0 \leq l \leq m-1$ 
because the saddle point $x_{s.p.}$ should be located within the interval 
$ 0 \leq x_{s.p.} \leq 1$ to give the finite contribution. 
Assuming that the boundary structure does not change in higher genus, 
we then get the expression
\bba 
  && \lambda \int^{m-1}_0 dl \biggl[ 
      \half \sum_{{\cal O}={\cal O}_1 \cup {\cal O}_2 \atop g=g_1 +g_2} 
         < \Tp_{m-1-l} ~{\cal O}_1 >_{g_1} < \Tp_l ~{\cal O}_2 >_{g_2} 
             \nonumber \\ 
  && \qquad\qquad\qquad 
        + \half < \Tp_{m-1-l} \Tp_l ~{\cal O} >_{g-1} \biggr] ~. 
\eea
The second term is a variant of the first term, which comes from 
the configuration that two surfaces are connected by a handle.   
The factor $\half$ in the first term is to correct the 
overcounting of the summation and that in the second term is to 
correct the double counting coming from the interchange of $\Tp_{m-1-l}$ 
and $\Tp_l$. 

  We next consider the $n=2$ case. We then have to evaluate the 
following quantity:
\bba
  && 
      -\fr{\lambda}{2} \int \fr{dl}{2\pi} 
         \int \fr{dh}{2\pi} < {\cal O}_1^{\pp} \biggl\{ 
         \int_{\e^{-\tau} \leq |z| \leq 1} d^2 z 
           {\bar \pd} W_{2,m} 
             \int_{|w| \leq |z|} d^2 w {\hat V}^+_k  +
            \\
  &&   \int_{\e^{-\tau} \leq |z| \leq 1} d^2 z {\hat V}^+_k 
             \int_{|w| \leq |z|} d^2 w {\bar \pd} W_{2,m}
          \biggr\} \fr{-2}{H} \e^{-\tau H} 
            | -h,-l>< h,l| ~{\cal O}_2^{\pp} > ~, 
              \nonumber
\eea
where $\Tp_k = \int d^2 z {\hat V}^+_k (z,\z)$. The primes on 
${\cal O}_{1,2}$ denote the exclusion of the operator $\Tp_k$.    
After carrying out the integration of $w$, we evaluate the $z$ 
and $h$ integral using the saddle point method. 
At $\tau \rightarrow \infty$ we obtain the contribution
\bb
   \lambda 2! ~k \int^{m-2+k}_0 dl 
     < {\cal O}_1^{\pp} \Tp_{m-2+k-l}>_{g_1} 
        <\Tp_l ~{\cal O}_2^{\pp} >_{g_2}~, 
\ee
where $g=g_1 +g_2$. There also is a variant of this contribution 
coming from the configuration where two surfaces are connected by 
a handle.

  As an another variant of (19) we furthermore obtain the boundary 
contribution with the triple product of amplitudes, 
\bb
   \lambda 2! \int^{m-2}_0 dl \int^{m-2-l}_0 d l^{\pp} 
     < \Tp_{m-2-l-l^{\pp}} {\cal O}_1 >_{g_1} 
        <\Tp_l ~{\cal O}_2 >_{g_2}
        < \Tp_{l^{\pp}} ~{\cal O}_3 >_{g_3}~,
\ee
where $g=g_1 +g_2 +g_3$. Noting the factorization property discussed 
before that the intermediate state becomes on-shell after integrating over 
the intermediate momentum, it is calculated by replacing the vertex 
operator ${\hat V}^+_k$ in (18) with the factrization formula 
$-\fr{\lambda}{2}\fr{2}{H_{l^{\pp}}} {\hat V}_{-l^{\pp}} 
< \Tp_{l^{\pp}} ~{\cal O}_2 >_{g_3}$, where 
$\fr{1}{H_{l^{\pp}}}= \int \fr{dh}{2\pi}(\half h^2 + \half l^{\pp 2})^{-1} 
= \fr{1}{l^{\pp}}$ and only the $N=0$ mode is considered. 

  The general $n$ formula is  
\bba
   &&  \lambda^{a-1} ~n! ~\biggl( \prod^{n+1-a}_{i=1} k_i \biggr) 
              \int \prod^a_{i=1} dl_i \theta(l_i) 
                   ~\delta \biggl( \sum^a_{i=1}l_i 
                        -\sum^{n+1-a}_{i=1} k_i +n-m \biggr)  
                  \nonumber  \\
   && \qquad\qquad  \times   < \Tp_{l_1} ~{\cal O}^{\pp}_1  >_{g_1} 
             < \Tp_{l_2} ~{\cal O}^{\pp}_2 >_{g_2} 
                   \cdots <\Tp_{l_a} ~{\cal O}^{\pp}_a >_{g_a} ~,
\eea
where $\sum^a_{i=1}g_i =g$ and $ a=1, \cdots ,n+1$. 
$\theta$ is the step function. The $a=1$ formula is nothing 
but the contribution of the OPE (11). In addition, as discussed in the 
cases of $n=1$, there are many variants of this expression  
coming from the boundary configurations that some of the surfaces are 
connected by handles.  
  
  The formulas with the vertex $\Tm_p$ are given by changing the chirality;  
$\Tp \rightarrow \Tm$ and $W_{n,m} \rightarrow -W_{m,n}$. 
Summarizing the boundary contributions, we can write out the Ward 
identities. For example, we get
\bba
   0  &=&  \fr{1}{\pi}\int {\bar \pd}
             < W_{2,1} ~\Tp_{k_1} \Tm_{p_1} \Tm_{p_2} >_g 
                 \nonumber \\
     &=& -x < \Tp_{k_1} \Tm_{p_1} \Tm_{p_2} \Tm_1 >_g 
        -p_1 <  \Tp_{k_1} \Tm_{p_1 +1} \Tm_{p_2} >_g
                \nonumber \\ 
     && -p_2 <  \Tp_{k_1} \Tm_{p_1} \Tm_{p_2 +1} >_g
        + 2! x k_1 < \Tp_{k_1 -1} \Tm_{p_1} \Tm_{p_2} >_g  
                 \nonumber  \\ 
     && -\fr{\lambda}{2} \int^1_0 dl 
           < \Tm_{1-l} \Tm_l \Tm_{p_1} \Tm_{p_2} \Tp_{k_1} >_{g-1} 
                        \\  
     && +\fr{\lambda}{2} 2! k_1 \int^{k_1 -1}_0 dl 
                 < \Tp_{k_1 -1 -l} \Tp_l \Tm_{p_1}\Tm_{p_2} >_{g-1} 
                      \nonumber \\ 
     && + \lambda 2! k_1 \sum^g_{h = 0} \int^{k_1 -1}_0 dl 
                     < \Tm_{p_1} \Tp_{k_1 -1-l} >_h 
                     < \Tp_l \Tm_{p_2} >_{g-h} ~, 
               \nonumber
\eea 
where the first term is given by the OPE with the tachyon background 
$\mu_B T_0$, where $\mu_B$ is replaced with the renormalized one 
$\mu =-x$. The second and the third terms are respectively given by 
the OPE with $\Tm_{p_1}$ and $\Tm_{p_2}$.   
The fourth term is given by the perturbed OPE with $\Tp_{k_1}$ and 
$\mu_B T_0$. The last three terms are just anomalous contributions 
coming from the boundary of moduli space. 

   This is not the end of the strory. As for sphere amplitudes 
the $W_{\infty}$ identities form a closed set among those  
and we can solve the identities recursively.  
We can obtain all types of sphere 
amplitudes of the normalized tachyons that exactly agree with the 
matrix model ones. 
For higher-genus cases, however, there is a problem. The $W_{n,m}$ 
identities with $n,m \leq 2$ such as (22) are consistent with the 
matrix model results. But, for the general $W_{n,m}$ identity with 
$n,m >2$, further boundary contributions are necessary in order that 
the solution exists.   
It is easily imagined that there are the contributions shown in  
fig.1. On the basis of this figure we can speculate a generalisation 
of the formula (21) as follows:
\bba
   &&  \lambda^{a-1+h} \int \prod^a_{i=1} dl_i \theta(l_i) 
          ~D^{+(h)}_a (-l_1,\cdots,-l_a,k_1,\cdots,k_{n+1-a-2h};n,m) 
                  \nonumber  \\
   && \qquad\qquad  \times   < \Tp_{l_1} ~{\cal O}^{\pp}_1  >_{g_1} 
             < \Tp_{l_2} ~{\cal O}^{\pp}_2 >_{g_2} 
                   \cdots <\Tp_{l_a} ~{\cal O}^{\pp}_a >_{g_a} ~,
\eea
where $\sum^a_{i=1}g_i =g-h$ and $ a=1, \cdots ,n+1-2h$. 
$h$ stands for the genus of the surface $\Sigma$ in fig.1.  
$-l_i$ represents the conjugate mode of $\Tp_{l_i}$. 
The $\Sigma$-part just gives the connectivity matrix $D^{+(h)}_a$ at 
$\tau \rightarrow \infty$. The $h=0$ formula is nothing but (21). 
The $h\neq 0$ contributions exist for $g \geq h$ and $ n \geq 2h$,  
where note that the $h=1$ formula would contibute in the 
$W_{2,1}$ identities, but it vanishes due to the Seiberg condition. 

\begin{figure}
\begin{center}
\leavevmode
\epsfbox{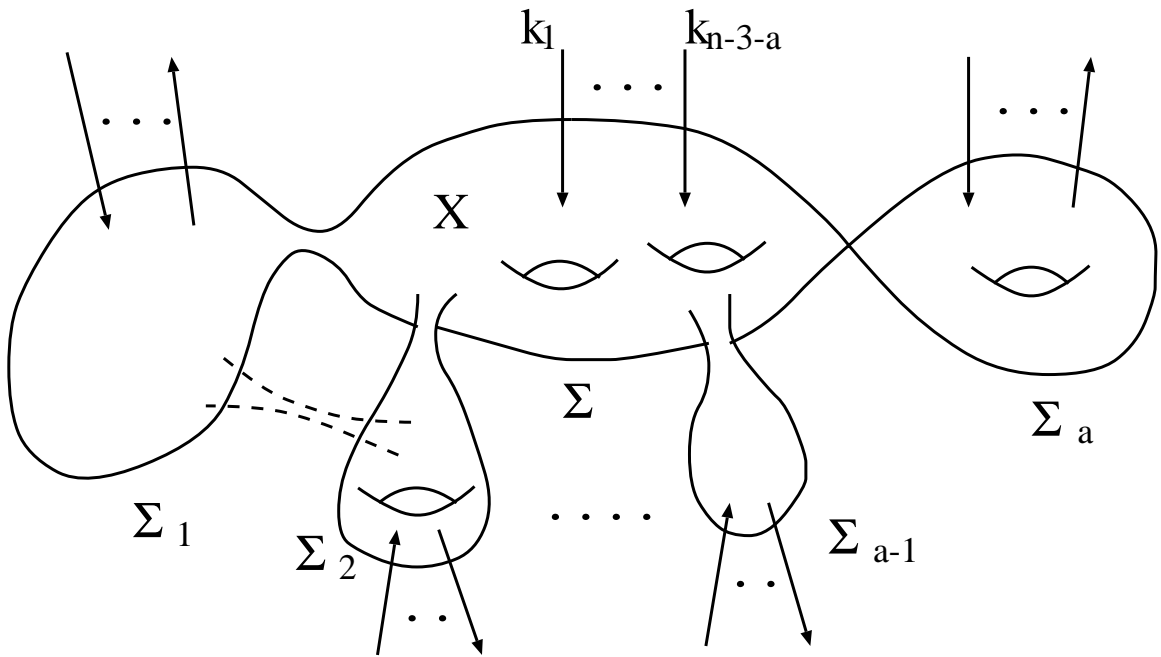}
\end{center}
\caption{The incoming and the outgoing arrows denote $\Tp_{k}$ and 
$\Tm_{p}$, respectively. The cross point is ${\bar \pd} W_{n,m}$. 
The degenerate point of the surface stands for $\fr{-2}{H}\e^{-\tau H}$. 
}
\label{fig:Fig.1} 
\end{figure}
 
  The direct calculation of the connectivity matrix $D^{\pm (h)}_a$ 
for $h \geq 1$ is very difficult.  
So we guess the forms. Recall that the discrete state 
$R_{n,m}$ is given by the OPE of the tachyon operators 
$ T^-_{n+1} \times T^+_{m+1} \sim R_{n,m} $. 
This suggests that we could replace the operator  ${\bar \pd}W_{n,m}$ 
with the two tachyons $\Tm_{n+1}$ and $\Tp_{m+1}$. 
Thus we identify the surface $\Sigma$ with the 
$1 \rightarrow n+2-2h$ amplitude of genus $h$.   
{}From this argument we guess the expressions of $D$ as follows:
\bba
   && D^{+(h)}_a (-l_1,\cdots,-l_a,k_1,\cdots,k_{n+1-a-2h};n,m) 
                     \\
   && \quad  = \fr{\lambda^{1-h}}{(n+1)(m+1)\prod^a_{i=1}(-l_i)}  
           ~{\hat S}^{(h)}_{m+1,-l_1,\cdots,-l_a,
                 k_1,\cdots,k_{n+1-a-2h} \rightarrow n+1} ~, 
                 \nonumber
\eea
where the $\S$-matrix formula is applied as if $-l_i$ were positive. 
The normalization is fixed by fitting the $h=0$ formula with (21). 
For example, using the result of the matrix model~\cite{k}, we obtain 
the genus-one expression
\bba
  && D^{+(1)}_a 
        = \delta ( \sum^a_{i=1} l_i -\sum^{n-1-a}_{i=1}k_i +n-m) 
            ~\fr{1}{24}~ n! ~\biggl( \prod^{n-1-a}_{i=1} k_i \biggr) 
                     \nonumber \\          
  && \qquad\qquad\qquad\quad \times
            \biggl( \sum^a_{i=1} l_i^2 + \sum^{n-1-a}_{i=1}k_i^2 
                        +(m+1)^2 -n-2 \biggr)  ~.
\eea
Once the connectivity matrix of genus one $D^{\pm (1)}_a$ is given, 
we can then obtain all genus amplitudes recursively.

\section{Summary}
 
  We studied the $W_{\infty}$ structure of 2D string theory in the 
continuum method. We derived recursion relations which connect 
different genus amplitudes. For sphere amplitudes we can solve 
the $W_{\infty}$ identities recursively and can obtain all types of 
amplitudes. For higher-genus cases we first checked that the $W_{n,m}$ 
identities with $n,m \leq 2$ are consistent with the matrix model. 
For general $W_{n,m}$, however, it is necessary to add the extra 
contributions (23) which are difficult to calculate directly. 
So we guessed the form from a simple argument and checked that the 
Ward identities are indeed closed and consistent with the matrix model. 
We explicitely verified the equivalence 
up to three genus by using the genus-one expression (25). 
In this way we conclude that the 
$\S$-matrix is equivalent with the matrix model amplitude in general 
genus. 

  Finally we comment on the work for the $c_M <1$ model~\cite{h}. 
In this case we consider the chiral theory that consists of only the 
positive (or the negative) 
tachyon states with the discrete momenta. Then the $W_{\infty}$ Ward 
identities result in the $W$-algebra constraints~\cite{fkn}.       

\vspace{5mm}
 
  I am grateful to Sumit R. Das for careful reading of the manuscript.

\end{document}